# Eigenspectra and Statistical Properties of the Klein-Gordon Equation with Cornell Potential: Unequal Mixings of Scalar and Time-Like Vector Potentials

F. Tajik[1], Z. Sharifi[1], M. Eshghi[2,*], M. Hamzavi[1], M. Bigdeli[1], S.M. Ikhdair[3,4]

[1]*Department of Physics, University of Zanjan, Zanjan, Iran*

[2]*Young Researchers and Elite Club, Central Tehran Branch, Islamic Azad University, Tehran, Iran*

[3] *Department of Physics, Faculty of Science, An-Najah National University, Nablus, Palestine*

[4] *Department of Electrical Engineering, Near East University, Nicosia, Northern Cyprus, Mersin 10, Turkey*

## Abstract

The D-dimensional Klein-Gordon (KG) wave equation with unequal scalar and time-like vector Cornell interactions is solved by the Laplace transform method. In fact, we obtained the bound state energy eigenvalues of the spinless relativistic heavy quarkonium systems under such potentials. Further, the stationary states are calculated due to the good behavior of wave functions at the origin and at infinity. The statistical properties of this model are also investigated. Our results are found to be of great importance in particle physics.

## I. Introduction

Relativistic wave equations such as the Dirac and the Klein–Gordon (KG) equations have obtained much interest from many theoretical physicists in different fields of physics [1, 2]. The particle dynamics in relativistic quantum mechanics is explained by using KG and Dirac equations.

On the other hand, the most interesting choices for the investigated potentials are probably the spherically symmetric ones because of their wide applications in several areas of physics including nuclear and particle physics. For example, solutions of the wave equations are extremely significant particularly in describing the nuclear shell structure [3−8]. So far a variety of analytical methods are expanded and utilized in solving these wave equations. As an example, see; the Supersymmetry (SUSY) technique [9-11], the Nikiforov-Uvarov method (NU) [12-14], the asymptotic iteration method (AIM) [15-19], the variational [20], the hypervirial perturbation method [21], the shifted large 1/N expansion (SE), the modified shifted 1/N expansion (MSE) [22], the exact quantization rule (EQR) [23-25], the Laplace transform (LP) [26-28], the perturbative formalism [29, 30], the polynomial solution [31], the wave function ansatz method [32], the group theory [33-34], the path integral approach [35-38], $L^2$ approach [39] and the point canonical transformation (PCT) [40-43].

* ***Corresponding author Email:*** eshgi54@gmail.com

The LT method is extensively used to solve the quantum wave equations. It absolutely was used during the initial years of quantum mechanics by Schrödinger when discussing the radial eigenfunctions of the hydrogen atom [44] and later Enlefield approached the Schrödinger equation with the Coulomb, harmonic oscillator, exponential and Yamauchi potentials [45]. The hydrogen atom was also investigated using the LT method [46]. In the last years, the $1/x$ [47], Morse [48], D-dimensional harmonic oscillator [49], 3D pseudoharmonic and Mie-type [50], one-dimensional harmonic oscillator [51], the Mie-type [26, 27], the pseudo-harmonic oscillatory ring-shaped [28], and the Makarov [52] potentials were solved by means of LT method.

Nowadays, for spin zero particles, the KG equation is regarded as the relativistic form of the Schrödinger equation. In the other word, among various nonrelativistic and relativistic wave equations, the KG equation has been receiving much theoretical and phenomenological attention as it allows us to investigate spin-zero particles. The KG equation has been used to mathematical physics such as nonlinear optics, solid-state physics, and quantum field theory. As an example, Grave *et al* have considered three examples of vibrating systems whose mathematical explanations lead to the KG equation [53]. Also, recently, there have been many investigated on KG equation with various kinds of potentials by using various methods to describe the corresponding relativistic physical systems [54-67]. The idea of this work is to research and investigate the bound state energy eigenvalues and the wave functions of the D-dimensional KG equation with unequal mixings of the vector-scalar Cornell potentials by means of the LT approach. Further, the energy states are mainly used to obtain the partition function that is essentially used to study the essential statistical properties like the thermal quantities of the Cornell model.

This paper is organized as follows. In Section 2, we study the bound state energy eigenvalues of the spinless relativistic heavy quarkonium systems under unequal mixings of scalar and vector Cornell potentials using the LT technique. In brief, we also give in brief a review to the LT method in this section. We obtain the exact bound state energy eigenvalues and the corresponding wave functions of the D-dimensional KG equation with Cornell potential. In Section 3, we study the statistical properties and the thermal quantities such as the Helmholtz free energy, the mean energy, the entropy, and the specific heat. Finally, we give some of our conclusions in Section 4.

## 2. Bound States Energy

In spherical coordinates, the KG equation with time-like vector $V(r)$ and scalar $S(r)$ potentials can be written as follows (in units of $\hbar = c = 1$)

$$\left[\Delta_D + (E_{nl} - V(r))^2 - (M + S(r))^2\right]\Psi_{nlm}(r,\Omega_D) = 0, \tag{1}$$

with

$$\Delta_D = \nabla_D^2 = r^{1-D}\frac{\partial}{\partial r}(r^{D-1}\frac{\partial}{\partial r}) - \frac{\Lambda_D^2(\Omega_D)}{r^2}, \tag{2}$$

where $\Lambda_D^2(\Omega_D)$ is the hyperangular momentum operator [37, 43] reads as below

$$\Lambda_D^2 = -\sum_{i,j=1}^{D}\Lambda_{ij}^2, \Lambda_{ij} = x_i\frac{\partial}{\partial x_j} - x_j\frac{\partial}{\partial x_i}, \tag{3}$$

for all Cartesian components $x_i$ of the D-dimensional vector $(x_1, x_2, ..., x_D)$. The eigenvalues $\Lambda_D^2(\Omega_D)$ are written by [9]

$$\Lambda_D^2(\Omega_D) Y_l^m(\Omega_D) = \frac{l(l+D-2)}{r^2} Y_l^m(\Omega_D),\ D > 1 \tag{4}$$

where $Y_l^m(\Omega_D)$ is the hyperspherical harmonics. For the case $D = 3,$ we have a familiar form of (4) as $\Lambda_{D=3}^2(\Omega_{D=3}) Y_l^m(\Omega_{D=3}) = \frac{l(l+1)}{r^2} Y_l^m(\theta,\phi).$ Now, by employing the separation of variables using the following wave function $\Psi_{nlm}(r,\Omega_D) = R_{nl}(r) Y_l^m(\Omega_D),$ Eq. (1) simplifies as below:

$$\left[\frac{d^2}{dr^2}+\frac{D-1}{r}\frac{d}{dr}-(M^2-E_{nl}^2)+V^2(r)\right.$$
$$\left.-S^2(r)-2(E_{nl}V(r)+MS(r))-\frac{l(l+D-2)}{r^2}\right]R_{nl}(r)=0, \tag{5}$$

with unequal mixture of scalar and time-like vector Cornell potentials taking the simple form:

$$V(r)=-\frac{a_v}{r}+b_v r, \tag{6a}$$

$$S(r)=-\frac{a_s}{r}+b_s r. \tag{6b}$$

After substituting the above potentials, then Eq. (5) turn to as follows:

$$\left[\frac{d^2}{dr^2}+\frac{D-1}{r}\frac{d}{dr}+\frac{a_v^2-a_s^2-l(l+D-2)}{r^2}+\frac{2(Ea_v+Ma_s)}{r}\right.$$
$$\left.+(b_v^2-b_s^2)r^2-2(Eb_v+Mb_s)r-\varepsilon^2-2(a_v b_v-a_s b_s)\right]R_{nl}(r)=0, \tag{7}$$

where $\varepsilon^2=M^2-E_{nl}^2$. In the limit when $r\to\infty$, the wave function satisfies the following differential equation:

$$\left(\frac{d^2}{dr^2}-\beta^2 r^2\right)R_{nl}(r)=0, \tag{8}$$

where we have considered the fact $(b_v^2-b_s^2)r^2 \gg -2(Eb_v+Mb_s)r$, for larger $r$ and assigned

$$\beta=\sqrt{b_s^2-b_v^2},\ b_s^2>b_v^2. \tag{9}$$

The solution of the Eq. (8) in the limiting case when $r^2 \gg 1$ is easy, with the solution as $R_{nl}(r)\propto e^{\frac{-\beta r^2}{2}}$.

Now, we seek a complete general solution to Eq. (7) which needs to be bounded, well defined at the origin and finite at the boundaries. Let us look for a solution in its general form as

$$R_{nl}(r)=r^k e^{\frac{-\beta r^2}{2}} f(r),\ k>0, \tag{10}$$

The term $r^k$ assures us that the solution at $r=0$ is bounded. Now, we try to find the function $f(r)$. The first and second order derivatives of Eq. (10) will be as below

$$\frac{dR_{nl}}{dr}=r^k e^{\frac{-\beta r^2}{2}}\left\{\frac{df(r)}{dr}+\frac{k}{r}f(r)-\beta r f(r)\right\}, \tag{11}$$

and

$$\frac{d^2R_{nl}}{dr^2}=r^k e^{\frac{-\beta r^2}{2}}\left\{\frac{d^2f(r)}{dr^2}+\left(\frac{2k}{r}-2\beta r\right)\frac{df(r)}{dr}+\left(\frac{k(k-1)}{r^2}-2\beta k-\beta+\beta^2r^2\right)f(r)\right\}, \qquad (12)$$

respectively.

Inserting them back into Eq. (7), we can obtain the following equation as

$$\begin{aligned} &rf''(r)+\left\{(2k+D-1)-2\beta r^2\right\}f'(r)+\\ &\left\{\frac{k(k-1)+k(D-1)-l(l+D-2)+(a_v^2-a_s^2)}{r}\right. \\ &\left.-2(Eb_v+Mb_s)r^2+2(Ea_v+Ma_s)+(-2\beta k-\beta D-\varepsilon^2-2(a_vb_v-a_sb_s)r\right\}f(r)=0, \end{aligned} \qquad (13)$$

where the prime over $f(r)$ denotes the derivative with respect to $r$. To apply the LT method on the above differential equation we must impose the following parametric condition

$$k(k-1)+k(D-1)-l(l+D-2)+(a_v^2-a_s^2)=0. \qquad (14)$$

Under the above condition, Eq. (13) reduces into a rather more simplified form as

$$\begin{aligned} &rf''(r)+\left\{(2k+D-1)-2\beta r^2)\right\}f'(r)+\left\{-2(Eb_v+Mb_s)r^2+2(Ea_v+Ma_s)\right. \\ &\left.+(-2\beta k-\beta D-\varepsilon^2-2(a_vb_v-a_sb_s)r\right\}f(r)=0. \end{aligned} \qquad (15)$$

Let us now implement the main basic ideas of the LT method so that we can solve the above differential equation [49-51, 71-73]. It is simply an integral transform and comprehensively useful in both physics and engineering. Recently, LT method has been utilized by many authors to solve the Schrodinger equation for various potential models [44, 46, 48, 51, 71-73]. The advantage of this method is that it converts the second-order differential equation into a first-order differential equation whose solutions may be obtained simply. We trace Refs. [74, 75] to display the basic ideas of the LT method as follows:

$$\phi(s)=L\{f(t)\}=\int_0^\infty e^{-st}f(t)dt, \qquad (16a)$$

$$L\{f^{(n)}(t)\}=s^nL\{f(t)\}-\sum_{k=0}^{n-1}s^{n-1-k}f^{(k)}(0), \qquad (16b)$$

$$L\{t^nf(t)\}=(-1)^n\phi^{(n)}(s), \qquad (16c)$$

$$L\{tf(t)\} = -\frac{d\phi(s)}{ds}, \tag{16d}$$

$$L\{tf''(t)\} = -2s\phi(s) - s^2\frac{d\phi(s)}{ds} + f(0);\ f(0) = 0, \tag{16e}$$

$$L\{tf'(t)\} = s\phi(s), \tag{16f}$$

where the superscript (n) denotes the *n*-th derivative with respect to $t$ for $f^{(n)}(t)$, and with respect to $s$ for $\phi^{(n)}(s)$. Furthermore, taking the following boundary condition $f(0) = 0$, Eq. (15) reduces to the following equation

$$(s+\alpha)\frac{d^2\phi(s)}{ds^2} + \left(\frac{1}{2\beta}s^2 + \lambda\right)\frac{d\phi(s)}{ds} + \left(\gamma s - \frac{(Ea_v + Ma_s)}{\beta}\right)\phi(s) = 0, \tag{17}$$

where the following identifications are used as below

$$\alpha = \frac{(Ea_v + Ma_s)}{\beta};\ \lambda = -\frac{\varepsilon^2}{2\beta} + \left(-k - \frac{D}{2} + 2\right) - \frac{(a_v b_v - a_s b_s)}{\beta};\ \gamma = \frac{3 - 2k - D}{2\beta}, \tag{18}$$

with $s = -\alpha$ is the singular point of the Eq.(17).

If there is some constant $\sigma \in R$ such that $\left|e^{-\sigma t} f(t)\right| \le M$ for sufficiently large *t*, then the integral in Eq. (16a) will exist for Re $s > \sigma$. The LT may fail to exist because of a sufficiently strong singularity in the function f(t) as $t \to 0$. In particular, we have

$$L\left[\frac{t^\alpha}{\Gamma(\alpha+1)}\right] = \frac{1}{s^{\alpha+1}},\ \alpha > -1. \tag{19}$$

Notice that if $s_0$ be as the singular point, then the LT behaves as $s \to s_0$

$$\phi(s) = \frac{1}{(s - s_0)^\nu}. \tag{20}$$

Then for the limiting case when $t \to \infty$, we have

$$f(t) = \frac{1}{\Gamma(\nu)} t^{\nu-1} e^{s_0 t}, \tag{21}$$

where $\Gamma(\nu)$ is the gamma function. So applying the fact of Eq. (20) the solution of Eq. (17) may be taken

$$\phi(s) = \frac{C_{nlD}}{(s+\alpha)^{n+1}},\ s_0 = -\alpha,\ \nu = \mathrm{n}+1,\ n = 0,1,2,3,... \tag{22}$$

Plugging Eq. (22) in Eq. (17) we can obtain the following three identity relations,

$$\gamma = \frac{n+1}{2\beta}, \tag{23}$$

$$\alpha = \frac{Ea_v + Ma_s}{\beta},\ \beta = \sqrt{b_s^2 - b_v^2},\ b_s \geq b_v \tag{24}$$

$$(n+1)(n+2) - (n+1)\lambda - \frac{(Ea_v + Ma_s)\alpha}{\beta} = 0. \tag{25}$$

Applying the set of Eqs. (23)-(25) together with Eq. (18), the energy eigenvalues will be as

$$E^2 - \frac{2Mb_v}{b_s}E - \left(1 - \frac{b_v^2}{b_s^2}\right)\left[M^2 + 2\left(a_v b_v - a_s b_s + (k + n + \frac{D}{2})\sqrt{b_s^2 - b_v^2}\right)\right] - M^2 = 0. \tag{26}$$

where $k$ can be found from the Eq. (14) as below

$$k = (2-D) \pm \sqrt{(D-2)^2 - 4\left[(a_v^2 - a_s^2) - l(l+D-2)\right]}. \tag{27}$$

We calculate the numerical results of the energy levels for positive sign, (+) for the various dimensional at the different values in Table 1.

On the other hand, the undetermined wave function $f(r)$ may be obtain from the inverse LT such that $f(r) = L^{-1}\{\phi(s)\}$ [41, 42]. Hence, using the Eq. (21) with $v = n+1$ and $s_0 = -\alpha$ we can obtain

$$f(r) = \frac{C_{nlD}}{n!} r^n e^{-\alpha r}. \tag{28}$$

Finally, with the help of the Eq. (28), the wave function of this system in Eq. (10) can be simply determined as

$$R_{nlD}(r) = \frac{C_{nlD}}{n!} r^{k+n} \exp\left(-\sqrt{b_s^2 - b_v^2}\frac{r^2}{2} - \left(\frac{Ea_v + Ma_s}{\sqrt{b_s^2 - b_v^2}}\right)r\right). \tag{29}$$

Let us now obtain the normalization constant $C_{nlD}$ using the condition [68, 76] as below: $\int_0^\infty \left[R_{nlD}(r)\right]^2 r^{D-1} dr = 1.$

A fair approximation $r + \frac{\alpha}{2\beta} \approx r$ would make the integration much easier to evaluate. As a consequence, the use of the following integral formula

$$\int_0^\infty x^p e^{-Ax^q} dx = \frac{1}{q}\frac{\Gamma(h)}{A^h},\ p,q>0, \tag{30a}$$

where $h=\dfrac{p+q}{q}$ leads to

$$C_{nlD} = n!\sqrt{\frac{2(\beta)^{k+n+D/2}}{\Gamma(k+n+D/2)}}. \tag{30b}$$

## 3. Statistical Properties of the KG-Cornell Model

To investigate the statistical properties, we need to obtain the partition function of the system using [77]

$$Z(\beta) = \sum_{n=0}^{n} e^{-\beta(E_n - E_0)}, \tag{31}$$

where $\beta = 1/k_B T$, $k_B$ is the Boltzmann constant and $T$ is the temperature in Kelvin. We start by recasting Eq. (26) as follows

$$E_n = M\frac{b_v}{b_s} \pm M\left[\alpha + 4n\delta\right]^{\frac{1}{2}}, \tag{32}$$

with the identifications $\alpha = \dfrac{1}{2M^2}\sqrt{b_s^2 - b_v^2}\left(1 - \dfrac{b_v^2}{b_s^2}\right)$ and

$\delta = \dfrac{2}{M^2}\left(1-\dfrac{b_v^2}{b_s^2}\right)\left(a_v b_v - a_s b_s + \sqrt{b_s^2 - b_v^2}\left(k+\dfrac{D}{2}\right)\right)$, then using the Euler-McLaurin formula [77, 78],

$$\sum_{m=0}^{\infty} f(m) = \frac{1}{2}f(0) + \int_0^\infty f(x)dx - \sum_{i=1}^{\infty}\frac{1}{(2i)!}B_{2i} f^{(2i-1)}(0), \tag{33}$$

where $B_{2i}$ are the Bernoulli numbers, $B_2 = 1/6, B_4 = -1/30, \ldots$, taking up to $i=2$, the canonical partition function be obtained as follows

$$Z(\mu) = \frac{1}{2} + \frac{\mu}{2\alpha}\left(\mu + \sqrt{\delta}\right) + \frac{\alpha^3}{60\mu^3\sqrt{\delta^3}}\left(\left(\frac{\delta}{\alpha^3} - \frac{2}{\delta}\right)\mu^2 - \frac{2}{\sqrt{\delta}}\mu - 1\right), \tag{34}$$

where $\mu = \dfrac{1}{M\beta} = \dfrac{k_B T}{M}$.

As follows, we study thermal quantities such as the Helmholtz free energy, the mean energy, the entropy and the specific heat as below

$$F(\beta) = -\frac{1}{\beta}\ln Z(\beta),\ U(\beta) = -\frac{\partial}{\partial\beta}\ln Z(\beta), \tag{35a}$$

$$S(\beta) = k_B\beta^2\frac{\partial}{\partial\beta}F(\beta),\ C_v(\beta) = -k_B\beta^2\frac{\partial}{\partial\beta}U(\beta), \tag{35b}$$

It would be better to rewrite above quantities as a function of $\mu = \frac{1}{M\beta}$ in the following forms:

$$\bar{F}(\mu) = \frac{F}{M} = -\mu\ln Z(\mu),\ \bar{U}(\mu) = \frac{U}{M} = \mu^2\frac{\partial}{\partial\mu}\ln Z(\mu), \tag{36a}$$

$$\bar{S}(\mu) = \frac{S}{K_B} = \ln Z(\mu) + \mu\frac{\partial}{\partial\mu}\ln Z(\mu), \tag{36b}$$

$$\bar{C}_v(\mu) = \frac{C_v}{K_B} = 2\mu\frac{\partial}{\partial\mu}\ln Z(\mu) + \mu^2\frac{\partial^2}{\partial\mu^2}\ln Z(\mu), \tag{36c}$$

It is noticed that for the limiting case at high temperatures corresponding to $\beta \ll 1$, our results reduce to

$$Z(\mu) \sim \mu^2,\ U(\mu) \sim 2\mu,\ C \sim 2, \tag{37}$$

These results can be easily inspected from the above figures.

## 4. Numerical Results

The vibrational bound state energies of scalar and time-like vector Cornell potential have been calculated for various *n* and *l* quantum numbers in *D*-dimensions. The obtained energy values are showed in Table 1 for $n = 1, 2, 3, 4, 5$, $l = 0, 1, 2, 3, 4$ and $D = 1, 2, \ldots, 6$ in order to give a wider energy spectrum. We see that the energy levels are decreasing with increasing dimensional and angular momentum, but it is increasing with increasing quantum numbers.

We also presented the behavior of the thermodynamics properties as a function of the potential parameters. In fact, our results have been showed in figures 1, 2, 3 and 4 where we have plotted the variation of all thermal quantities versus the dimensionless parameter $\mu$ for the Cornell potential. In Fig. 1, we see that the Helmholtz free energy decreases exponentially with increasing of $\mu$, but, in Fig. 2, the mean energy increases linearly. Also, the entropy decrease with increasing the parameter $\mu$ in Fig. 3. In fact, the beginning, it decreases with the increasing $\mu$, then it reaches a minimum at around $\mu = 0$, after the entropy begins to increase exponentially. In Fig. 4, the specific heat capacity increases exponentially with increasing the dimensionless parameter $\mu$.

## 5. Conclusions

We have investigated the solution of the D-dimensional KG wave equation with unequal couplings of scalar and time-like vector Cornell potentials by using the LT method since the Cornell potential plays an important role in the study of quarkonium and baryonic systems. Therefore, we have solved the KG equation with the Cornell potential and obtained the exact bound state energy eigenvalues and the corresponding wave functions. The statistical properties of this model have been investigated and the thermal quantities such as the Helmholtz free energy, the mean energy, the entropy and the specific heat have been plotted as shown in Figures 1- 4. The behavior of the thermal quantities versus the dimensionless parameter $\mu = 1/M\beta$ obtained at high temperature shows that our results reduce to the limit as a function of $\bar{\mu}$ as follows $Z(\mu) \approx \mu^2, U(\mu) \approx 2\mu, C \approx 2.$

## Competing Interests

The authors declare that there is no conflict of interest regarding the publication of this paper.

## References

1. A. W. Thomas, W. Weise, "*Structure of the Nucleon*", Wiley-VCH, Berlin, 2001.
2. B. Thaller, "*The Dirac Equation*", Springer-Verlag, New York, 1992.
3. C.-S. Jia, P. Guo, and X.-L. Peng, *J. Phys. A: Math*. **39** (2006) 7737.
4. R. Sever, C. Tezcan, M. Aktas, and O. Yesiltas, *J. Math. Chem*. **43** (2007) 845.
5. S. M. Ikhdair and M. Hamzavi, *Chin. Phys. B* **21** (2012) 110302.
6. M. R. Pahlavani and S. M. Motevalli, *Int. J. Theor. Phys.* **48** (2009) 1622.
7. S. M. Ikhdair and R. Sever, *Int. J. Mod. Phys. C* **20** (2008) 361.
8. S. M. Ikhdair and R. Sever, *Int. J. Mod. Phys. E* **17** (2008) 1107.
9. F. Cooper, A. Khare, and U. Sukhatme, *Phys. Rep.* **251** (1995) 267.

10. D. A. Morales, Chem. *Phys. Lett.* **394** (2004) 68.
11. S.-H. Dong, *Factorization Method in Quantum Mechanics* (Springer, The Netherlands), 2007.
12. M. Eshghi and S.M. Ikhdair, *Math. Meth. Appl. Sci.* **37** (2014) 2829.
13. M. Eshghi, H. Mehraban and S.M. Ikhdair *Pramana-J. Phys.* **88** (2017) 73.
14. A. F. Nikiforov and V. B. Uvarov, *Special Functions of Mathematical Physics* (Birkh¨auser, Basel), 1988.
15. M. Eshghi and H. Mehraban, *Chin. J. Phys.* **54**(4) (2012) 533.
16. M. Eshghi, *Acta Sci. Tech.* **34**(2) (2012) 207.
17. O. Bayrak and I. Boztosun, *J. Phys. A: Math. Gen.* **39** (2006) 6955.
18. H. Ciftci, R. L. Hall, and N. Saad, *Phys. Lett. A* **340** (2005) 388.
19. M. Eshghi, R. Sever and S.M. Ikhdair, *Eur. Phys. J. Plus* **131** (2016) 223.
20. E. D. Filho and R. M. Ricotta, *Phys. Lett. A* **269** (2000) 269.
21. J. B. Killingbeck, A. Grosjean, and G. Jolicard, *J. Chem. Phys.* **116** (2002) 447.
22. M. Bag, M. M. Panja, and R. Dutt, *Phys. Rev.* A **46** (1992) 6059.
23. Z. Q. Ma and B. -W. Xu, *Europhys. Lett.* **69** (2005) 685.
24. W. -C. Qiang and S. -H. Dong, *Eur. Phys. Lett*. **89** (2010) 10003.
25. W. -C. Qiang, G. -H. Sun, and S. -H. Dong, *Ann. Phys.* (*Berlin*) **524** (2012) 360.
26. M. Eshghi and S.M. Ikhdair, *Z. Naturforsch*. **69a** (2014) 111.
27. M. Eshghi and S.M. Ikhdair *Chin. Phys. B* **23**(12) (2014) 120304.
28. M. Eshghi, *Chin. Phys. Lett*. **29**(11) (2012) 110304.
29. B. Gonul, K. Koksal, and E. Bakir, *Phys. Scr.* **73** (2006) 279.
30. S. M. Ikhdair and R. Sever, *Int. J. Mod. Phys. A* **21** (2006) 6465.
31. S. M. Ikhdair and R. Sever, *J. Mol. Struct.*: (THEOCHEM) **806** (2007) 155.
32. S. H. Dong, *Int. J. Theor. Phys*. **40** (2001) 569.
33. S. H. Dong, *Phys. Scr.* **65** (2002) 289.
34. Y. Alhassid, F. Gursey, and F. Iachello, *Ann. Phys.* **148** (1983) 346.
35. R. Feynman and A. Hibbs, *Quantum Mechanics and Path Integrals* (McGraw-Hi11, New York) 1965.
36. L. S. Schulman, *Techniques and Applications of Path Integration* (Wiley, New York), 1981.
37. I. H. Duru and H. Kleinert, *Phys. Lett. B* **84** (1979) 185.

38. N. K. Pak and I. Sokmen, *Phys. Rev. A* **30** (1984) 1629.
39. M. Eshghi and S. Arbabi Moghaddam *Math. Meth. Appl. Sci.* **38** (2015) 5124.
40. R. De, R. Dutt, and U. Sukhatme, *J. Phys. A: Mat. Gen.* **25** (1992) L843.
41. R. Dutt, A. Khare, and Y. P. Varshni, *J. Phys. A* **28** (1995) L107.
42. C. -S. Jia, Y. -F. Diao, M. -Li, Q. -B. Yang, L. -T. Sun, and R. -Y. Huang, *J. Phys. A* **37** (2004) 11275.
43. A. D. Alhaidari, *Phys. Rev. Lett.* **87** (2001) 210405.
44. E. Schrodinger, *Ann. Phys.* **384** (1926) 361.
45. M. J. Englefield, *J. Aust. Math. Soc.* **8** (1968) 557.
46. R. A. Swainson and G. W. F. Drake, *J. Phys. A: Math. Gen.* **24** (1991) 79.
47. Y. Ran, L. Xue, S. Hu and R. K. Su, *J. Phys. A: Math. Gen.* **33** (2000) 9265.
48. G. Chen, *Phys. Lett. A* **326** (2004) 55.
49. C. Gang, *Chin. Phys.* **14** (2005) 1075.
50. A. Arda and R. Sever, *J. Math. Chem.* 5**0** (2012) 971.
51. D. R. M. Pimentel and A. S. de Castro, *Eur. J. Phys.* **34** (2013) 199.
52. M. Eshghi *Can. J. Phys.* **91** (2013) 71.
53. P. Grave and C. Gauthier, *Am. J. Phys*. **79** (2011) 447.
54. E. Olgar, R. Koc, and H. Tutunculer, *Chin. Phys. Lett.* **23**(3) (2006) 539.
55. C.-F. Hou, X.-D. Sun, Z.-X. Zhou, and Y. Li, *Acta Phys. Sin.* **48**(3) (1999) 385.
56. C. Y. Chen, C. L. Liu, F. L. Lu, and D. S. Sun, *Acta Phys. Sin.* **52** (2003) 1579.
57. Y. He, Z. Q. Cao, and Q. S. Shen, *Phys. Lett. A* **326**(5-6) (2004) 315.
58. G. Chen, Z.-D. Chen, and Z.-M. Lou, *Phys. Lett. A* **331**(6) (2004) 374.
59. W.-C. Qiang, *Chin. Phys.* **12**(10) (2003) 1054.
60. L.-Z. Yi, Y.-F. Diao, J.-Y. Liu, and C.-S. Jia, *Phys. Lett. A* **333**(3-4) (2004) 212.
61. C.-S. Jia, Y. Li, Y. Sun, J.-Y. Liu, and L.-T. Sun, *Phys. Lett. A* **311**(2-3) (2003) 115.
62. F. Dom´ ınguez-Adame, *Phys. Lett. A* **136**(4-5) (1989) 175.
63. Z.-Q. Ma, S.-H. Dong, X.-Y. Gu, J. A. Yu, and M. LozadaCassou, *Int. J. Mod. Phys. E* **13**(3) (2004) 597.
64. C.-Y. Chen, D.-S. Sun, and F.-L. Lu, *Acta Phys. Sin*. **55**(8) (2006) 3875.
65. W.-C. Qiang and S.-H. Dong, *Phys. Lett. A* **372**(27-28) (2008) 4789.
66. S.-S. Dong, S.-H. Dong, H. Bahlouli, and V.-B. Bezerra, *Int. J. Mod. Phys. E* **20** (2011) 55.

67. S.M. Ikhdair and R. Sever, *Ann. der Phys.* **16**(3) (2007) 218.
68. T. Dos, A. Arda, arXive:13085295v2 [math-ph], (2014).
69. A. Chatterjee, *Phys. Rep.* **186** (1990) 249.
70. S.-H. Dong, "*Wave Equations in Higher Dimensions*", Springer-Verlag, 2011.
71. A. D. Polyanin and A. V. Manzhirov, "*Handbook of Integral equations*", New york, Wash-Ington: CRC press 1998.
72. S. Ortakaya, *Chin. Phys. B* **21 (**2012) 070303.
73. M. Zarezadeh and M. K. Tavassoly, *Chin. Phys. C* **37** (2013) 043106.
74. M. R. Spiegel, *Schaum's Outline of Theory and Problems of Laplace Transforms*, (Schaum Publishing Co.) 1965.
75. G. Doetsch, "*Introduction to the Theory and Application of the Laplace Transformations*", (New York: Springer) 1974.
76. R. Kumar and F. Chand, *Phys. Scr.* **85** (2012) 055008.
77. M. H. Pacheco, R. R. Landim and C. A. S. Almeida, *Phys. Lett A* **311** (2003) 93.
78. M. H. Pacheco, R. V. Maluf, C. A. S. Almeida and R. R. Landim, *EPL* **311** (2014) 10005

**Table 1:** Relativistic Energy of the system for different states and dimensions with $M = 1,\ b_v = 0.002,$ $b_s = 2,\ a_v = 0.2,\ a_s = 6.$

| | $l$ | $E_{n,l}^{D}(D=1)$ | $E_{n,l}^{D}(D=2)$ | $E_{n,l}^{D}(D=3)$ | $E_{n,l}^{D}(D=4)$ | $E_{n,l}^{D}(D=5)$ | $E_{n,l}^{D}(D=6)$ |
|---|---|---|---|---|---|---|---|
| $n=1$ | 0 | 6.012<br>-6.010 | 5.829<br>-5.827 | 5.670<br>-5.668 | 5.536<br>-5.534 | 5.427<br>-5.425 | 5.346<br>-5.344 |
| $n=2$ | 0 | 6.336<br>-6.334 | 6.163<br>-6.161 | 6.012<br>-6.010 | 5.886<br>-5.884 | 5.784<br>-5.782 | 5.708<br>-5.706 |
| | 1 | 6.336<br>-6.334 | 6.108<br>-6.106 | 5.899<br>-5.897 | 5.712<br>-5.710 | 5.550<br>-5.548 | 5.414<br>-5.412 |
| $n=3$ | 0 | 6.644<br>-6.642 | 6.479<br>-6.477 | 6.336<br>-6.334 | 6.216<br>-6.214 | 6.120<br>-6.118 | 6.048<br>-6.046 |
| | 1 | 6.644<br>-6.642 | 6.427<br>-6.425 | 6.229<br>-6.227 | 6.052<br>-6.050 | 5.899<br>-5.897 | 5.771<br>-5.769 |
| | 2 | 6.542<br>-6.540 | 6.264<br>-6.262 | 5.998<br>-5.996 | 5.750<br>-5.748 | 5.522<br>-5.520 | 5.321<br>-5.319 |
| $n=4$ | 0 | 6.939<br>-6.937 | 6.781<br>-6.779 | 6.644<br>-6.642 | 6.530<br>-6.528 | 6.439<br>-6.437 | 6.370<br>-6.368 |
| | 1 | 6.939<br>-6.937 | 6.731<br>-6.729 | 6.542<br>-6.540 | 6.374<br>-6.372 | 6.229<br>-6.227 | 6.108<br>-6.106 |
| | 2 | 6.841<br>-6.839 | 6.575<br>-6.573 | 6.323<br>-6.321 | 6.088<br>-6.086 | 5.873<br>-5.871 | 5.684<br>-5.682 |
| | 3 | 6.632<br>-6.630 | 6.289<br>-6.287 | 5.947<br>-5.945 | 5.612<br>-5.610 | 5.289<br>-5.287 | 4.984<br>-4.982 |
| $n=5$ | 0 | 7.221<br>-7.219 | 7.070<br>-7.068 | 6.939<br>-6.937 | 6.830<br>-6.828 | 6.742<br>-6.740 | 6.677<br>-6.675 |
| | 1 | 7.221<br>-7.219 | 7.022<br>-7.020 | 6.841<br>-6.839 | 6.680<br>-6.678 | 6.542<br>-6.540 | 6.427<br>-6.425 |
| | 2 | 7.127<br>-7.125 | 6.873<br>-6.871 | 6.632<br>-6.630 | 6.408<br>-6.406 | 6.205<br>-6.203 | 6.026<br>-6.024 |
| | 3 | 6.927<br>-6.925 | 6.599<br>-6.597 | 6.275<br>-6.273 | 5.958<br>-5.956 | 5.654<br>-5.652 | 5.370<br>-5.368 |
| | 4 | 6.586<br>-6.584 | 6.144<br>-6.142 | 5.676<br>-5.674 | 5.177<br>-5.175 | 4.644<br>-4.642 | 4.071<br>-4.069 |

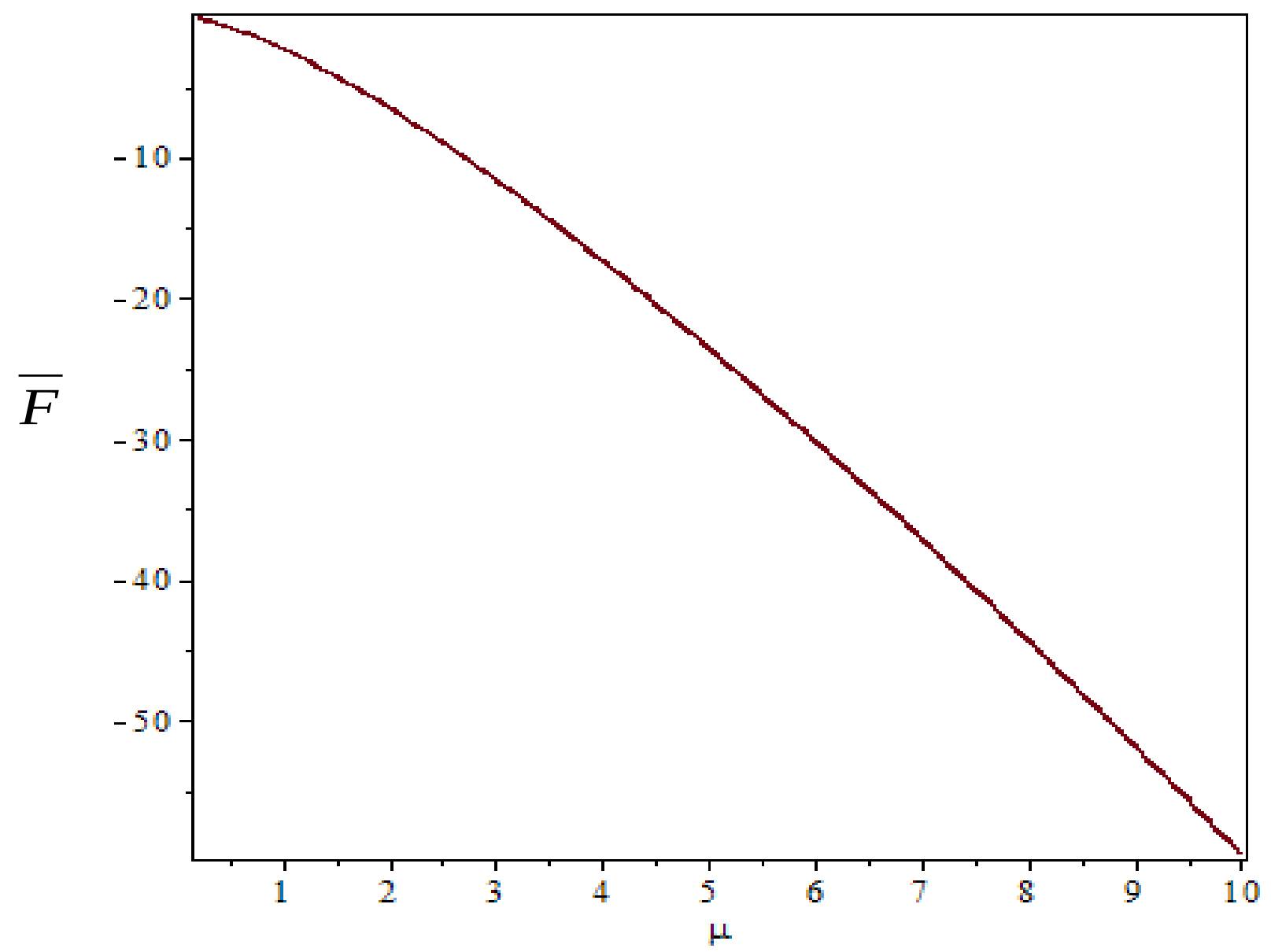

**Figure 1:** The free energy ($\overline{F}$) for the Cornell potential versus $\mu$

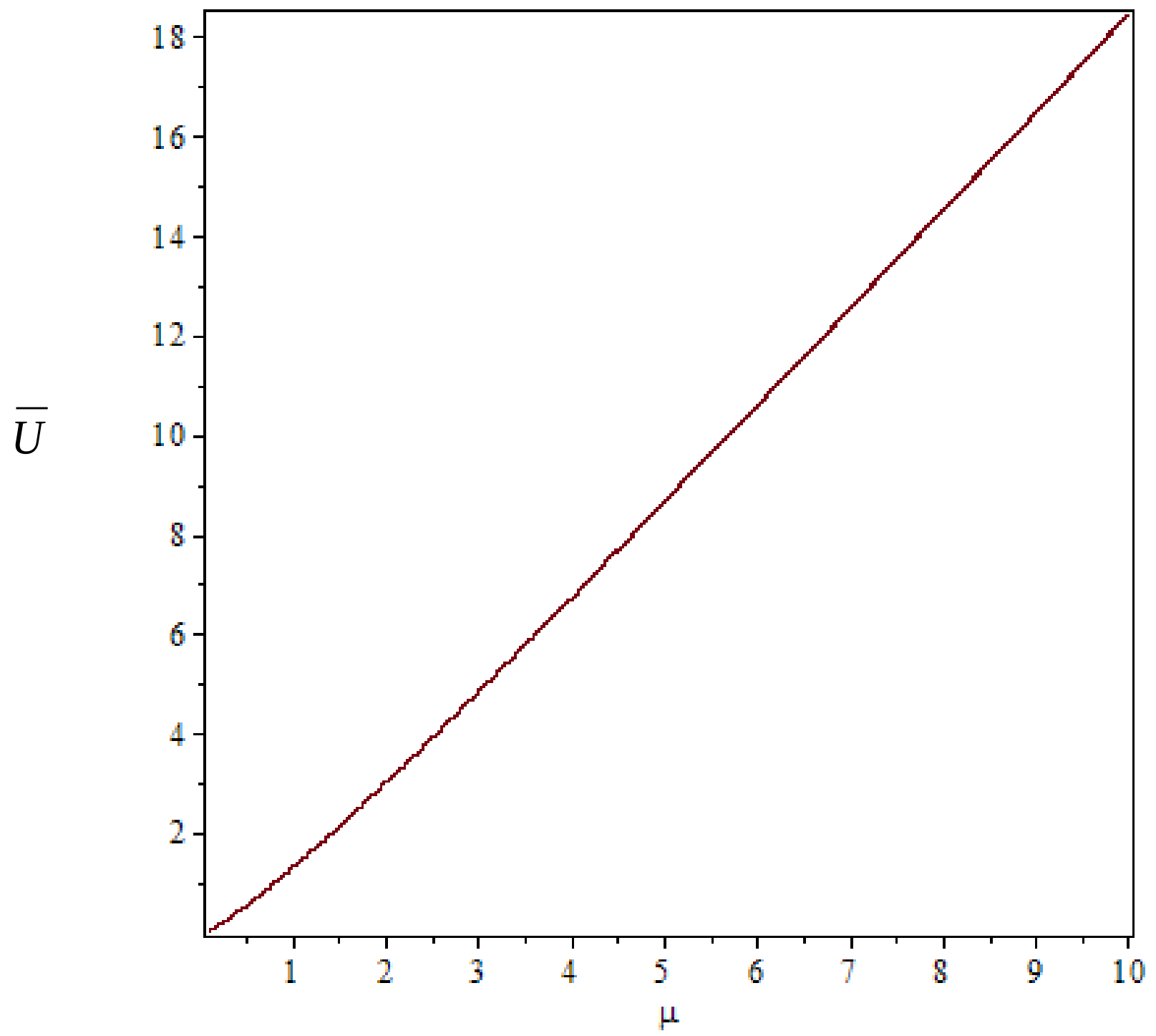

**Figure 2:** The mean energy ($\overline{U}$) for the Cornell potential versus $\mu$

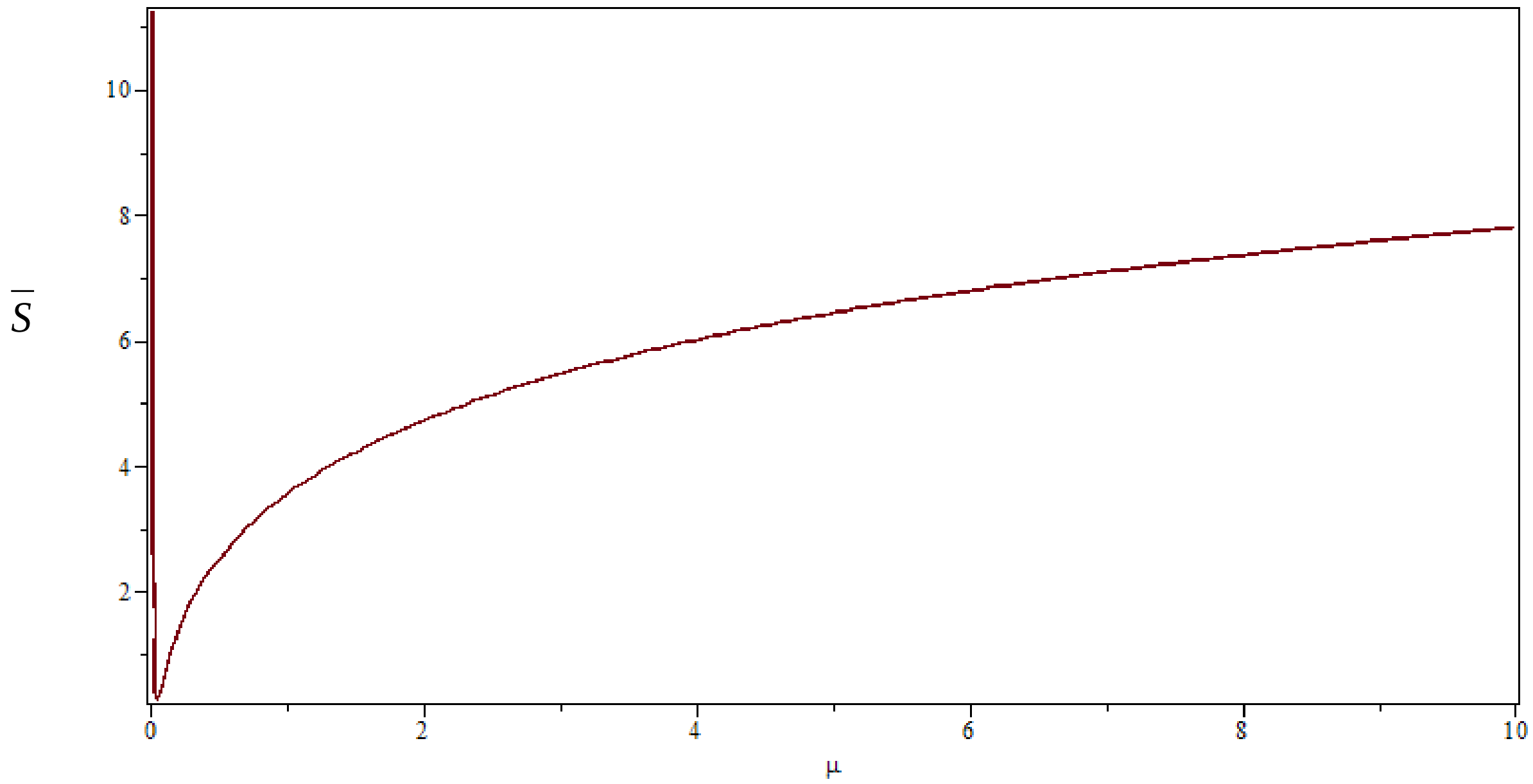

**Figure 3:** The entropy ( $\overline{S}$ ) for the Cornell potential versus $\mu$

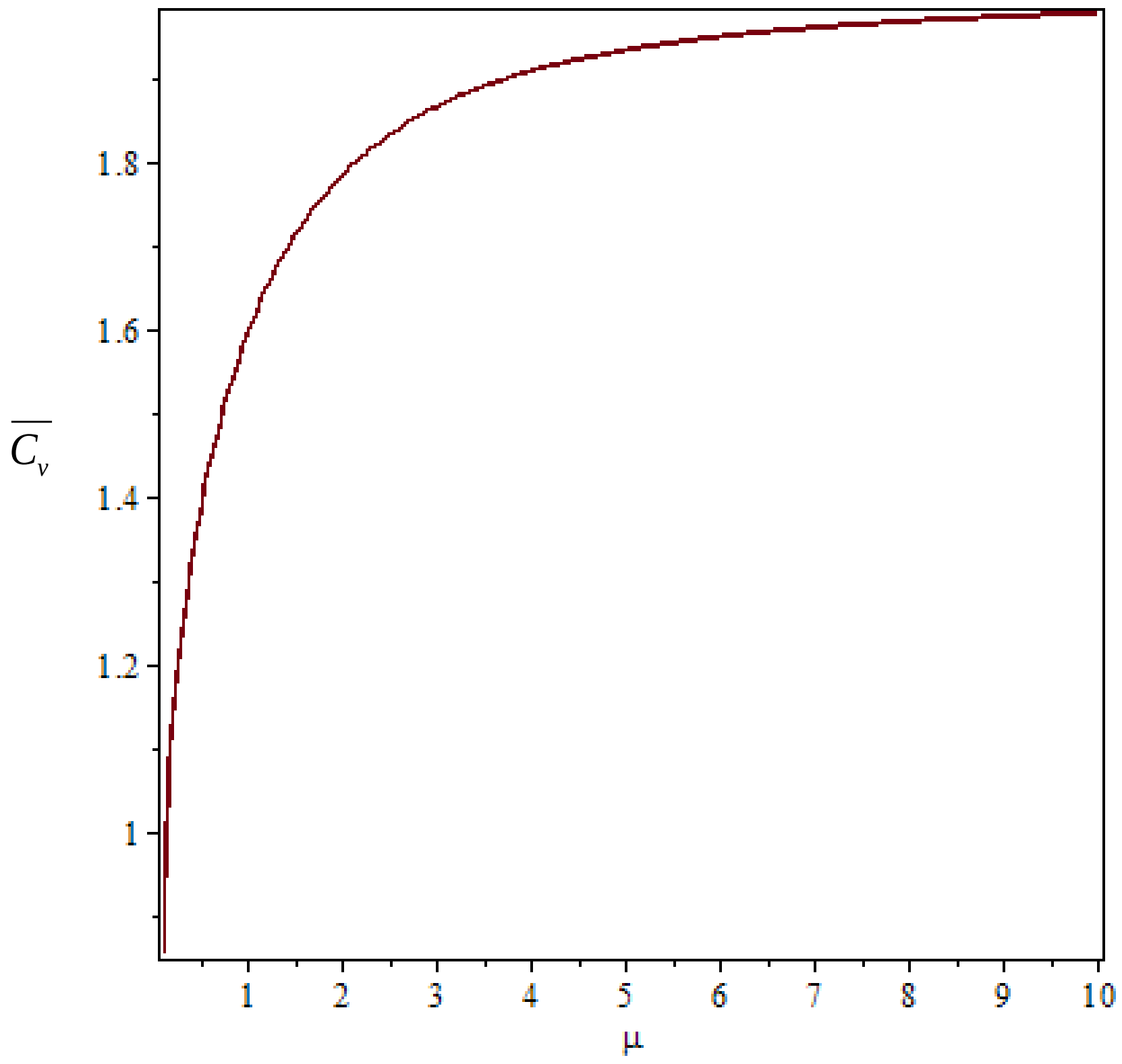

**Figure 4:** The specific heat ( $\overline{C_v}$ ) for the Cornell potential versus $\mu$